\documentstyle[12pt,epsfig]{article}
\newcommand{\tb}  {\mbox{$ \tan\beta~ $}} 

\newcommand{\besg}{$b  \to  X_s \gamma~ $}

\begin{document}


\begin{flushright}
IEKP-KA/2001-14 \\[3mm]
{\tt hep-ph/0106311}
\end{flushright}


\begin{center}
  {\large\bf A global fit to the anomalous magnetic moment, \besg  
   and Higgs limits in the constrained MSSM} \\[8mm]

  {\bf W. de Boer, M. Huber, C. Sander}
\\[2mm]
  {\it Institut f\"ur Experimentelle Kernphysik, University of Karlsruhe \\
       Postfach 6980, D-76128 Karlsruhe, Germany} \\[3mm]

  {\bf D.I. Kazakov} \\[2mm]

{\it Bogoliubov Laboratory of Theoretical Physics,
Joint Institute for Nuclear Research, \\
141 980 Dubna, Moscow Region, Russian Federation}

\end{center}

\begin{abstract}
{New data on the anomalous magnetic moment
of the  muon together with the  \besg decay rate 
 are considered  within the supergravity inspired
constrained minimal supersymmetric model.
We perform a  global statistical $\chi^2$ analysis of  these data
 and show that the allowed region of parameter space is bounded
from below by the Higgs limit, which depends on the trilinear
coupling and from above by the anomalous magnetic
moment $a_\mu$. The newest \besg data deviate 1.7 $\sigma$
from recent SM calculations and prefer a similar parameter
region as the 2.6 $\sigma$ deviation from $a_\mu$.}
\end{abstract}

\section{Introduction}

Recently  a new measurement of the anomalous magnetic moment
of the muon became available, which suggests a possible 2.6 standard
deviation from the Standard Model (SM) expectation\cite{BNL}: 
$\Delta a_\mu=a_\mu^{exp}-a_\mu^{th}=(43\pm 16)\cdot 10^{-10}$.
The theoretical prediction depends on the uncertainties in the
vacuum polarization and the light-by-light scattering, see e.g.
the discussion in~\cite{Y}.
However, even with a conservative estimate of the theoretical errors,
one has a positive difference $\Delta a_\mu$
of the order of the weak contribution to the anomalous magnetic moment, which
 opens a window for  "new physics".
 The most popular explanation is given in the framework
of SUSY theories~\cite{CM}-\cite{baer}, since
the contribution of superpartners to the
anomalous magnetic moment of the muon is 
  of the order of the weak contribution  and allows to explain 
the desired difference  $\Delta a_\mu$.
It requires the Higgs mixing parameter to be positive\cite{CN}
and the sparticles contributing to the chargino-sneutrino
$(\tilde{\chi}^\pm - \tilde{\nu}_\mu)$
and neutralino-smuon $(\tilde{\chi}^0 - \tilde{\mu})$ loop diagrams 
 to be relatively light\cite{CM}.

\begin{figure}[ht]\vspace*{-15mm}\begin{center}
\epsfig{file=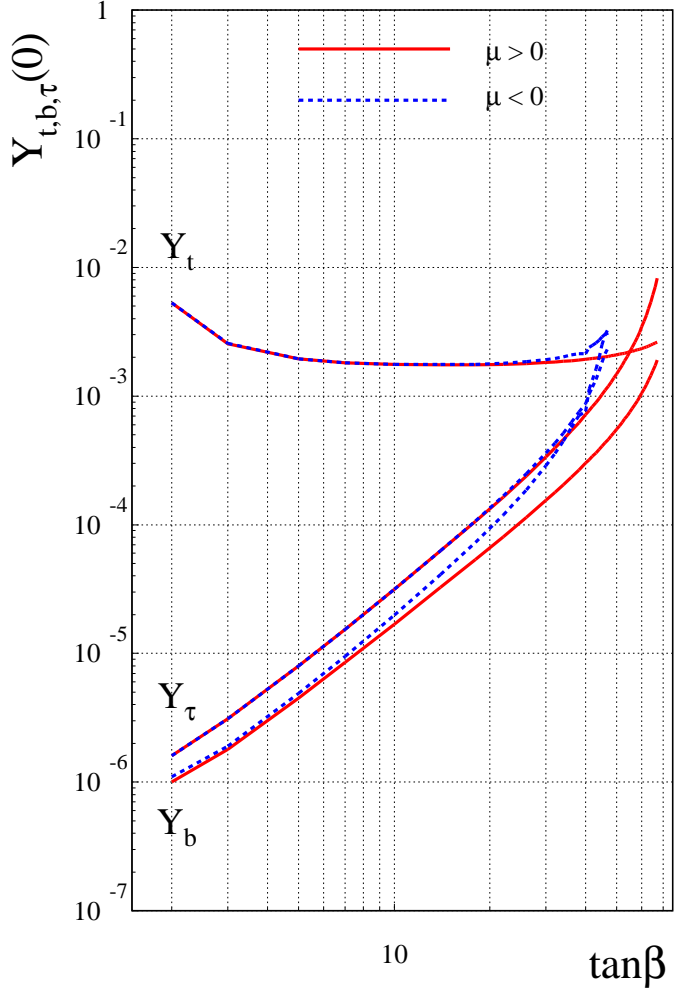,width=0.6\textwidth}
\caption[]{\label{f1} 
The dependence of the third generation  Yukawa couplings at the GUT scale
as function of $\tb$ for $\mu_0>0$ and $\mu_0<0$, obtained by fitting them to the
low energy masses of the top, bottom and tau mass. The results are for
a common mass $m_0=m_{1/2}=500$ GeV, but for different masses the curves look
very similar, except that the 'triple' unification point for $\mu_0<0$ shifts
between 42 and 48, if the common mass is shifted from 200 to 1000 GeV.}
\end{center}
\end{figure}

The positive sign of $\mu_0$ is also preferred by the branching ratio of the 
b-quark decaying radiatively into an s-quark - \besg - \cite{we}.
Last year the observed value of \besg was close to the SM expectation, so 
in this case the sparticles contributing to the chargino-squark
$(\tilde{\chi}^\pm - \tilde{q})$ and charged Higgs-squark
$(H^\pm - \tilde{q})$ loops
have to be rather heavy in order {\it not} to contribute to \besg.

However, it was recently suggested that in the theoretical calculation
one should use the running c-quark mass in the ratio $m_c/m_b$, which
reduces this ratio from 0.29 to 0.22 \cite{misiak}. The SM value for \besg increases from 
$(3.35\pm 0.30)\times 10^{-4}$ to $(3.73\pm 0.30)\times 10^{-4}$ 
in this case. This value is 1.7 $\sigma$ above the most recent 
world average of $(2.96\pm 0.46)\times 10^{-4}$, which is the average
from CLEO ($(2.85\pm 0.35_{stat}\pm 0.22_{sys})\times 10^{-4}$) \cite{CLEO}, 
ALEPH ($(3.11\pm 0.80_{stat}\pm 0.72_{sys})\times 10^{-4}$) \cite{ALEPH}  and 
BELLE ($(3.36\pm 0.53_{stat}\pm 0.42_{sys}(\pm^{0.50}_{0.54})_{model})\times 10^{-4}$) \cite{BELLE}.
For the error of the world average we added all errors in quadrature.

As will be shown, the small deviations from the SM for both $a_\mu$ and \besg
require now very similar mass spectra for the sparticles. 

%
In the Constrained Minimal Supersymmetric
Model (CMSSM) with supergravity mediated breaking terms all sparticles
masses are related by the usually assumed GUT scale boundary conditions of  a
 common mass $m_0$  for the  squarks and sleptons
 and a common mass $m_{1/2}$ for the gauginos.
The region of overlap in the GUT scale parameter space,
where both $a_\mu$ and \besg are within errors consistent with the data,
is most easily determined by a global statistical analysis, in which
the GUT scale parameters are constrained to the low energy data by
a $\chi^2$ minimization.

In this paper we present such an analysis within the CMSSM assuming
common scalar and gaugino masses and radiatively induced  electroweak
symmetry breaking.
We use the full NLO renormalization group equations to calculate
the low energy values of the gauge and Yukawa couplings and the one-loop
RGE equations for the sparticle masses with decoupling of the
contribution to the running of the coupling constants at threshold.
For the Higgs potential we use the full 1-loop contribution of all
particles and sparticles. For details we refer to previous
publications\cite{ZP,PL}.

In principle, one can also require $b-\tau$ Yukawa coupling unification,
which has a solution at low and high values of 
the ratio
of vacuum expectation values of the neutral components of the
two Higgs doublets, denoted  $\tan\beta=\langle H_2^0\rangle / 
\langle H_1^0 \rangle$\cite{ZP,PL}.
From Fig. \ref{f1} one observes that if the third generation
Yukawa couplings at the
GUT scale are constrained by the low energy top, bottom and tau masses,
they become equal for $\mu<0$ at $\tb\approx 40$, while for
$\mu>0$ they never become equal, although the difference
between the Yukawa couplings is less than a factor three.
Since $\mu>0$ is required by $\Delta a_\mu>0$ (see below),
we do not insist on Yukawa
coupling unification and consider $\tb$ to be a free parameter,
except for the fact that the present Higgs limit of 113.5 GeV
from LEP\cite{newhiggs}
requires $\tb>4.3$ in the CMSSM\cite{we}.

\begin{figure}[ht]\vspace*{-15mm}\begin{center}
\epsfig{file=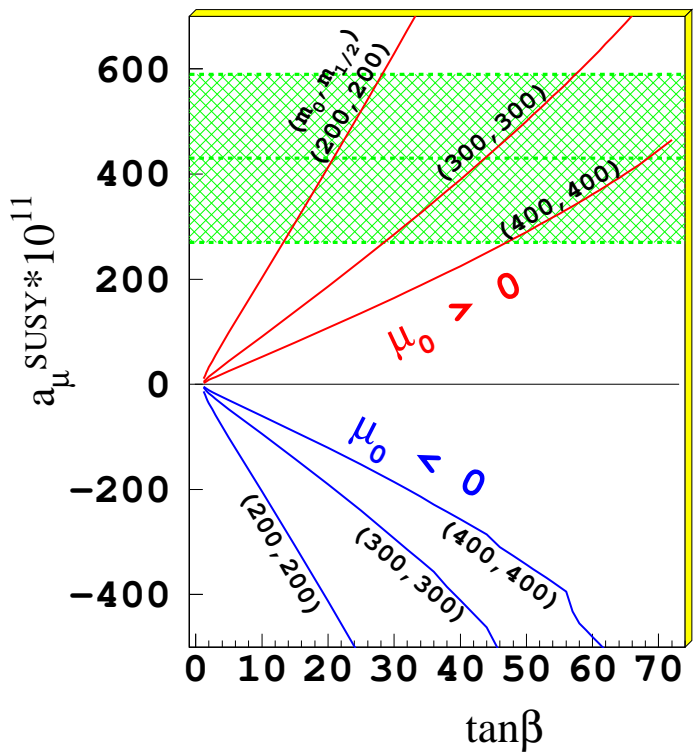,width=0.6\textwidth}
\caption[]{\label{f2} 
The dependence of  $a_\mu^{SUSY}$ versus $\tan\beta$  for various
values of the SUSY breaking parameters $m_0$ and $m_{1/2}$.
The horizontal band shows the discrepancy between the experimental data and
the SM estimate.
Good agreement with the data is only achieved at large $\tb$
and for light sparticles. Clearly, the fit allows only the positive sign of 
$\mu$.}\end{center}
\end{figure}
\begin{figure}[ht]
\begin{center}
\epsfig{file=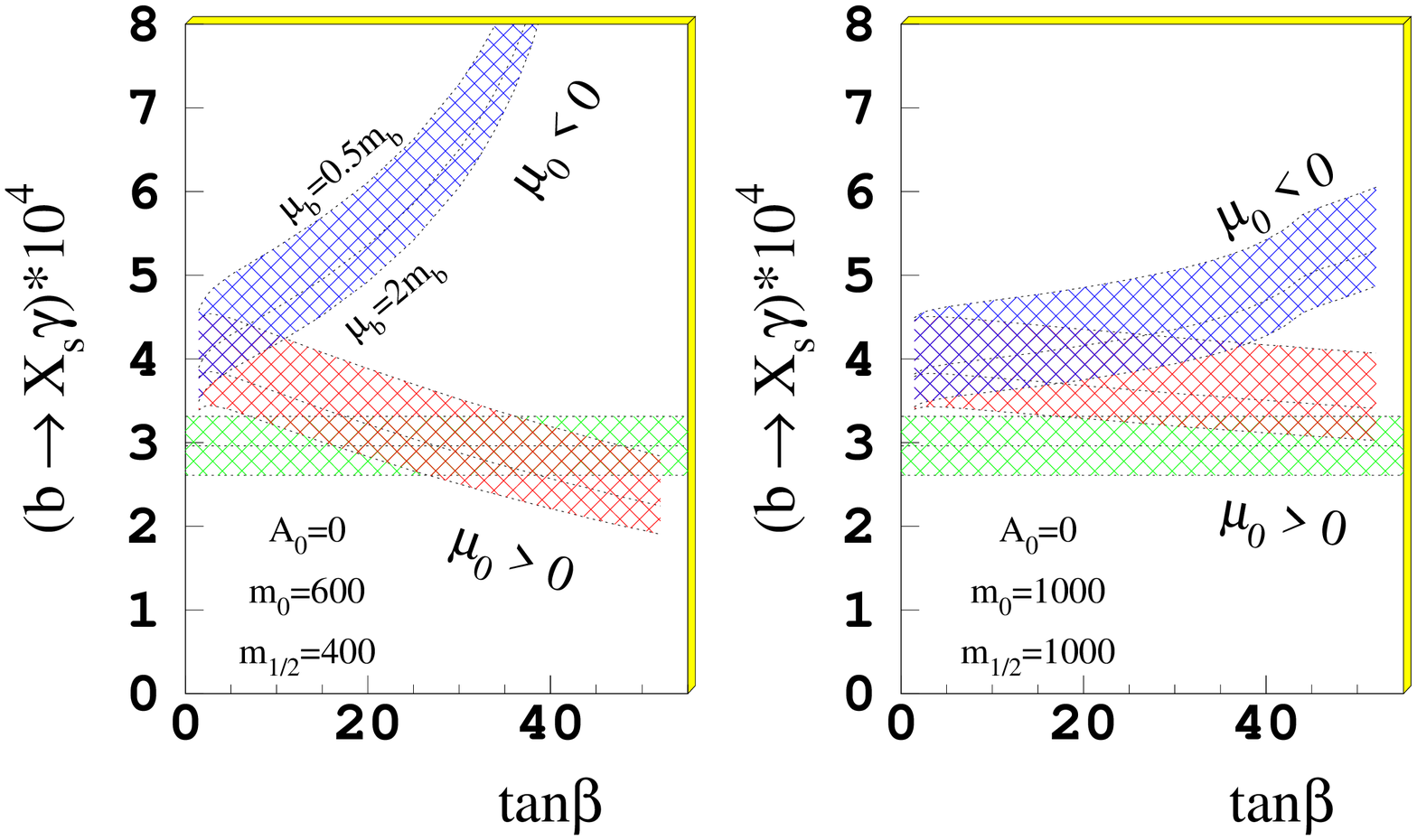,width=0.82\textwidth} 
\epsfig{file=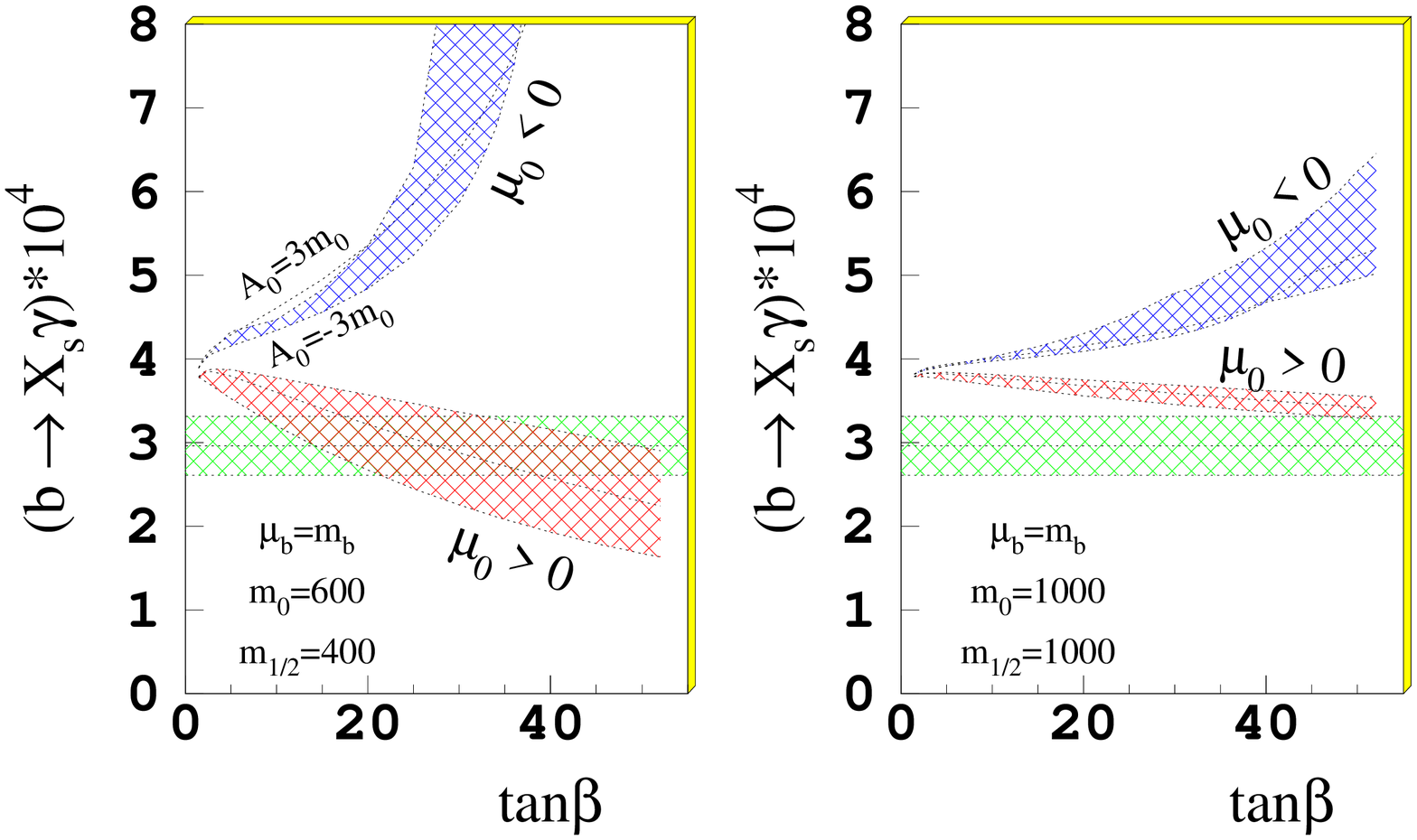,width=0.82\textwidth} 
\caption[]{\label{f3} 
The upper picture shows the dependence of the \besg \ rate on $\tan\beta$ 
 for $A_0=0$ and   m$_0$ = 600 (1000) GeV, m$_{1/2}$ = 400
(1000)GeV at the left (right).
For each value of $\tan\beta$ a fit was made to bring the predicted \besg 
rate (curved bands) as close as possible to the data (horizontal bands).
The width of the predicted values shows
the renormalization scale uncertainty from a scale variation
between 0.5m$_b$ and 2m$_b$.\\
The bottom picture shows the same dependence but for a fixed
renormalization scale of 1m$_b$. The width of the band is given
by the variation of $A_0$ between -3$m_0$ and 3$m_0$.}
\end{center}
\end{figure}
\begin{figure}[htb]
\epsfig{file=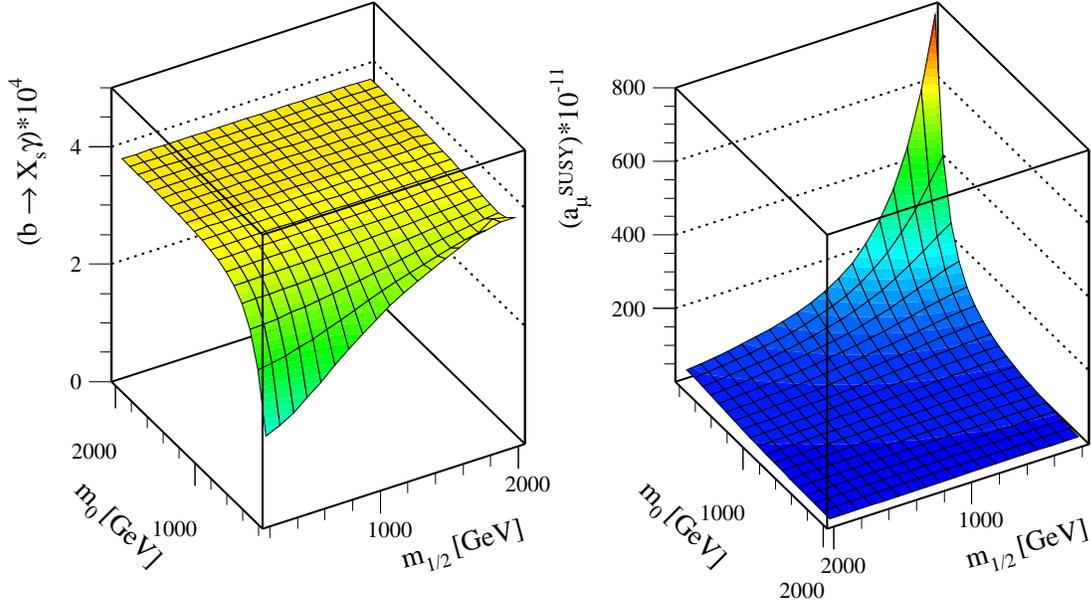,width=\textwidth}
\caption[]{\label{f4} 
The values of $b\to X_s\gamma$ and $a_\mu^{SUSY}$ in the $m_0,m_{1/2}$
plane for positive $\mu$ and $\tb=35$ to be compared with experimental data 
$b\to X_s\gamma = (2.96\pm 0.46) \cdot 10^{-4}$ and 
$a_\mu^{SUSY}=(43\pm 16)\cdot 10^{-10}$. One can see
that both $b\to X_s\gamma$ and $a_\mu^{SUSY}$ prefer relatively light sparticles.}
\end{figure}
\begin{figure}[htb]
\epsfig{file=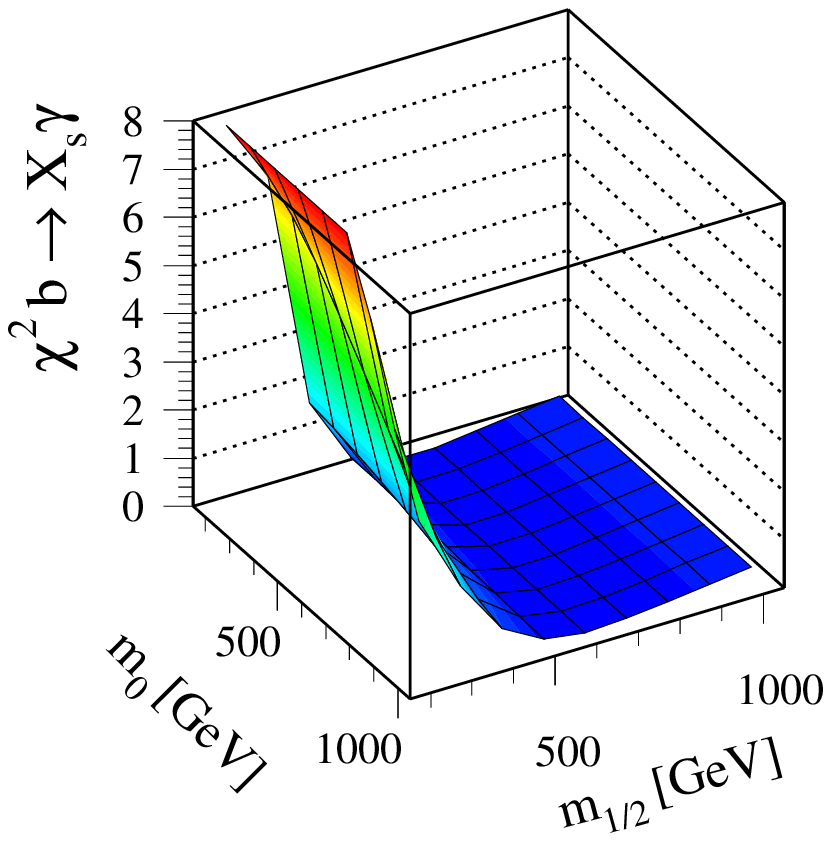,width=5cm}%
\epsfig{file=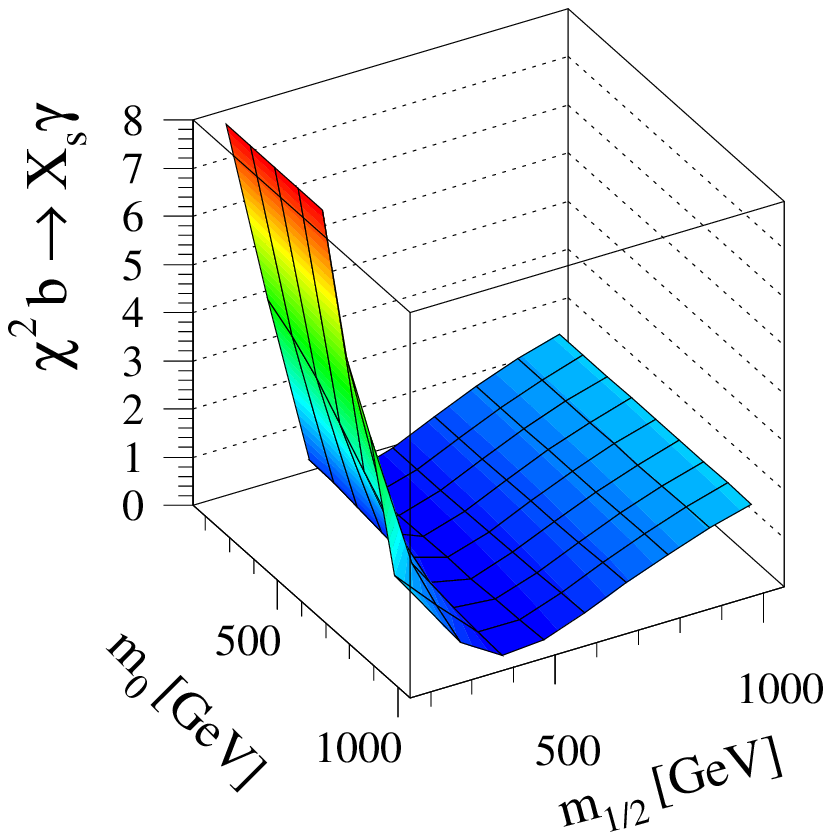,width=5cm}%
\epsfig{file=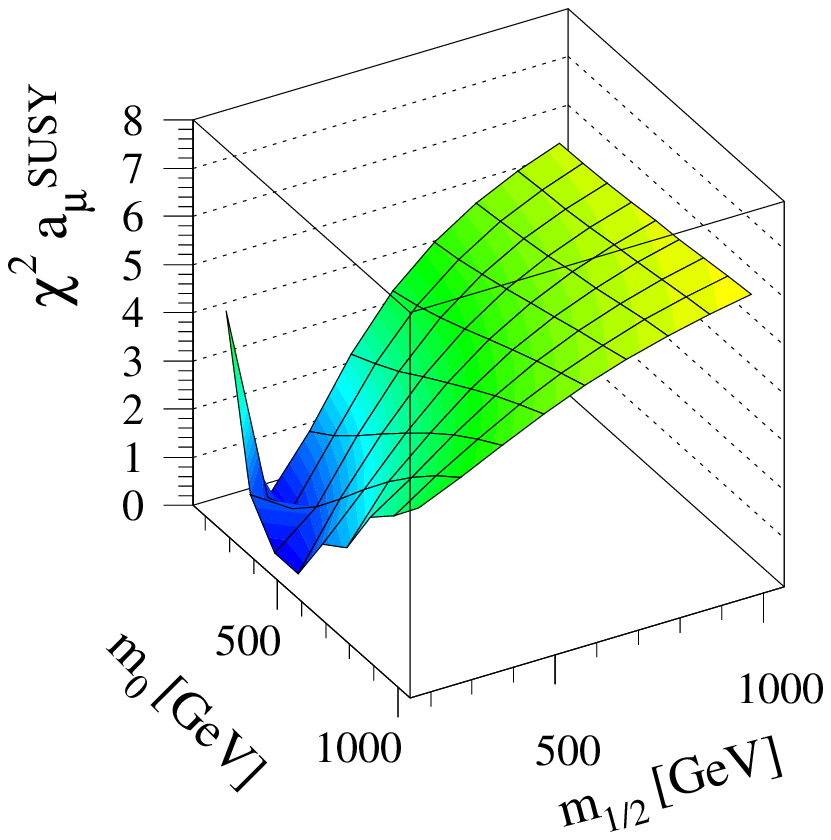,width=5cm}%
\caption[]{\label{f5} 
The individual contributions to $\chi^2$ from
$b\to X_s\gamma$ and $a_\mu$ in the  $m_0,m_{1/2}$ plane for $\tb=35$, $\mu>0$
and $A_0=0$. On the left handside we show the old contribution from \besg, as
calculated with $m_c/m_b=0.29$, which has the lowest $\chi^2$ for
heavy supersymmetric particles. In the middle the contribution from \besg for  $m_c/m_b=0.22$
is shown, which now has a minimum for intermediate masses. The $\chi^2$ contribution from $a_{\mu}$
is shown on the right handside, which clearly prefers light sparticles.
}
\end{figure}
\begin{figure}[htb]\vspace*{-5mm}
\epsfig{file=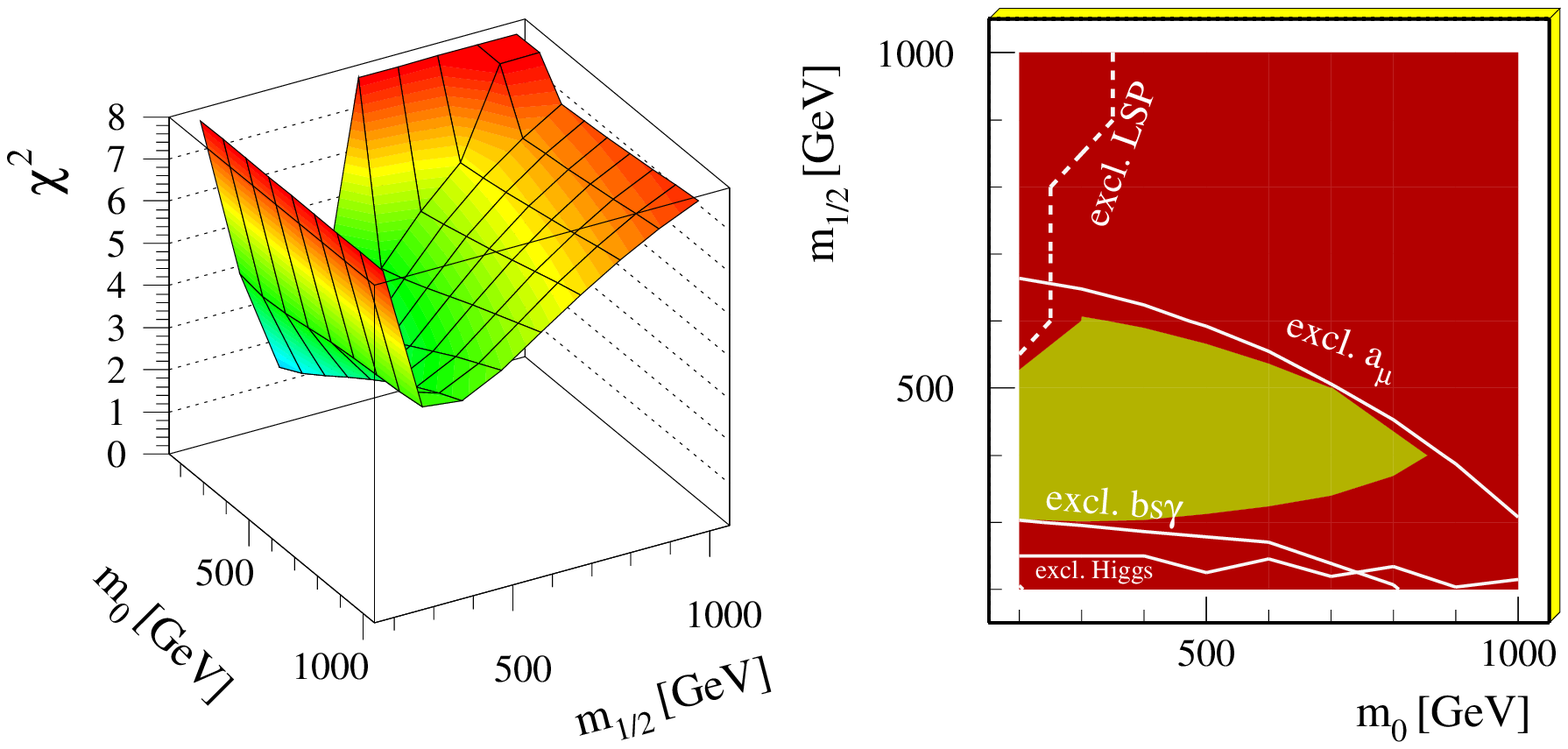,width=0.9\textwidth}
\epsfig{file=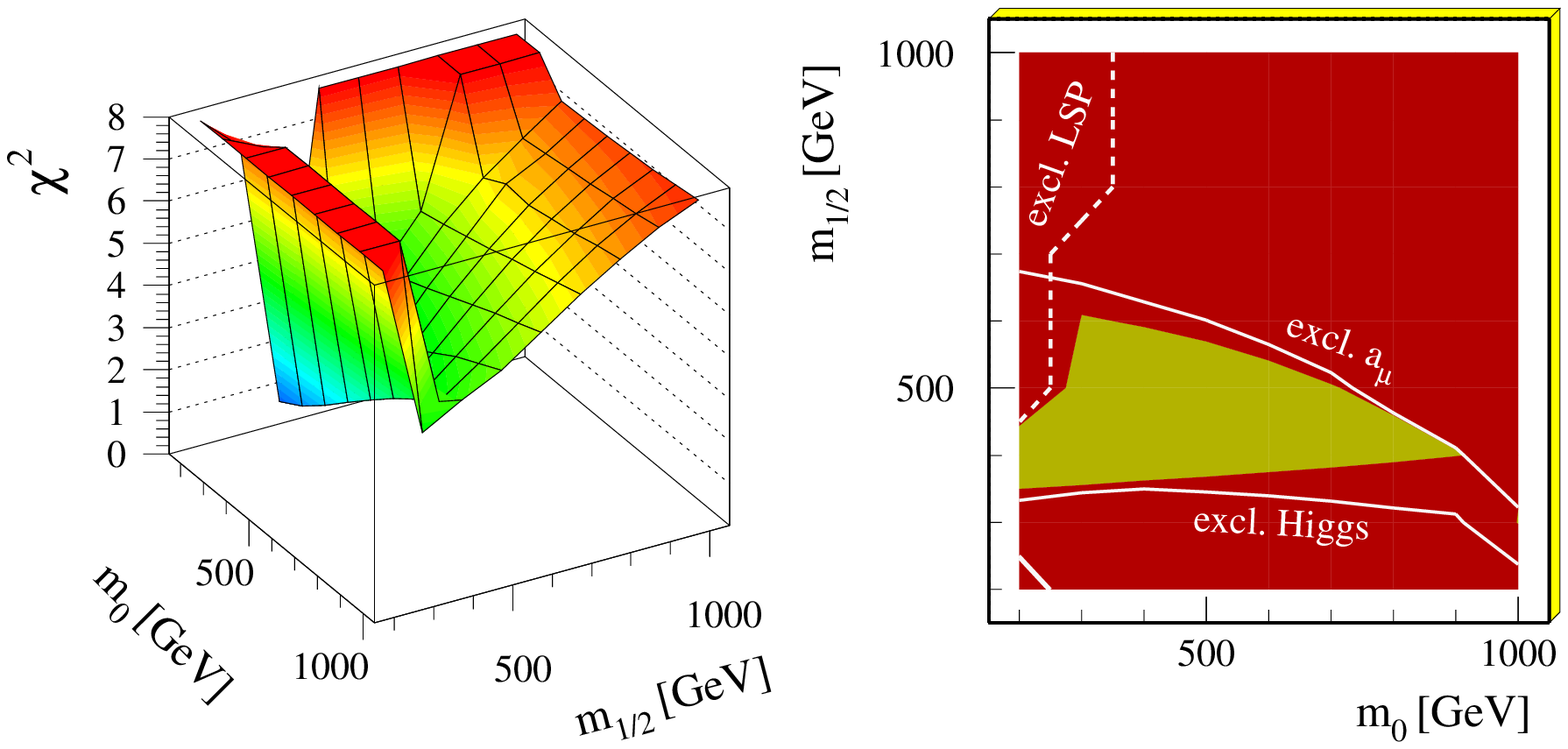,width=0.9\textwidth}
\caption[]{\label{f6} 
The upper part shows  the  $\chi^2$ contribution (left) and its projection
(right)  in the   $m_0, m_{1/2}$ plane 
  for $A_0=0$ and \tb= 35. 
The light shaded area is the region, where the combined $\chi^2$ is below
4. The regions outside this shaded region are excluded at 95\% C.L..
The white lines  correspond to the  "two-sigma" contours,
i.e. $\chi^2=4$ for that particular contribution.
The lower row shows the same for the fit, where $A_0$ was left free,
in which case $A_0\approx 3m_0$ (its maximum allowed value in our fit)
is preferred in the region where the stop mixing is important, 
i.e. regions where the left and right handed stops are not very heavy compared
with the top mass.
One observes that with $A_0$ as a free parameter the Higgs limit becomes
the most important lower bound on the SUSY sparticles, while for the no-scale
models with $A_0=0$ (top) the \besg rate determines mainly the lower bound.
%
}
\end{figure}
\begin{figure}[htb]\begin{center}
\epsfig{file=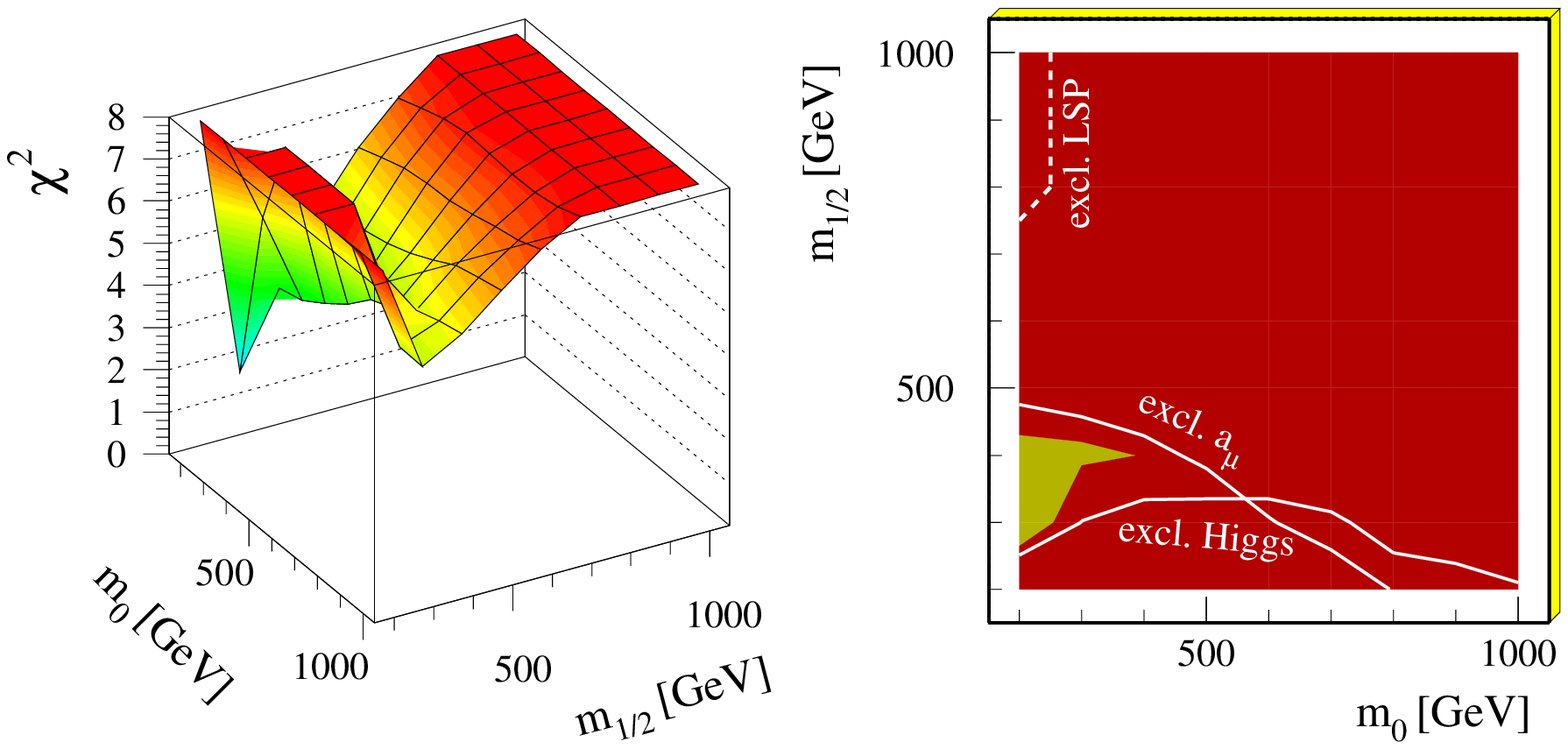,width=0.85\textwidth}
\epsfig{file=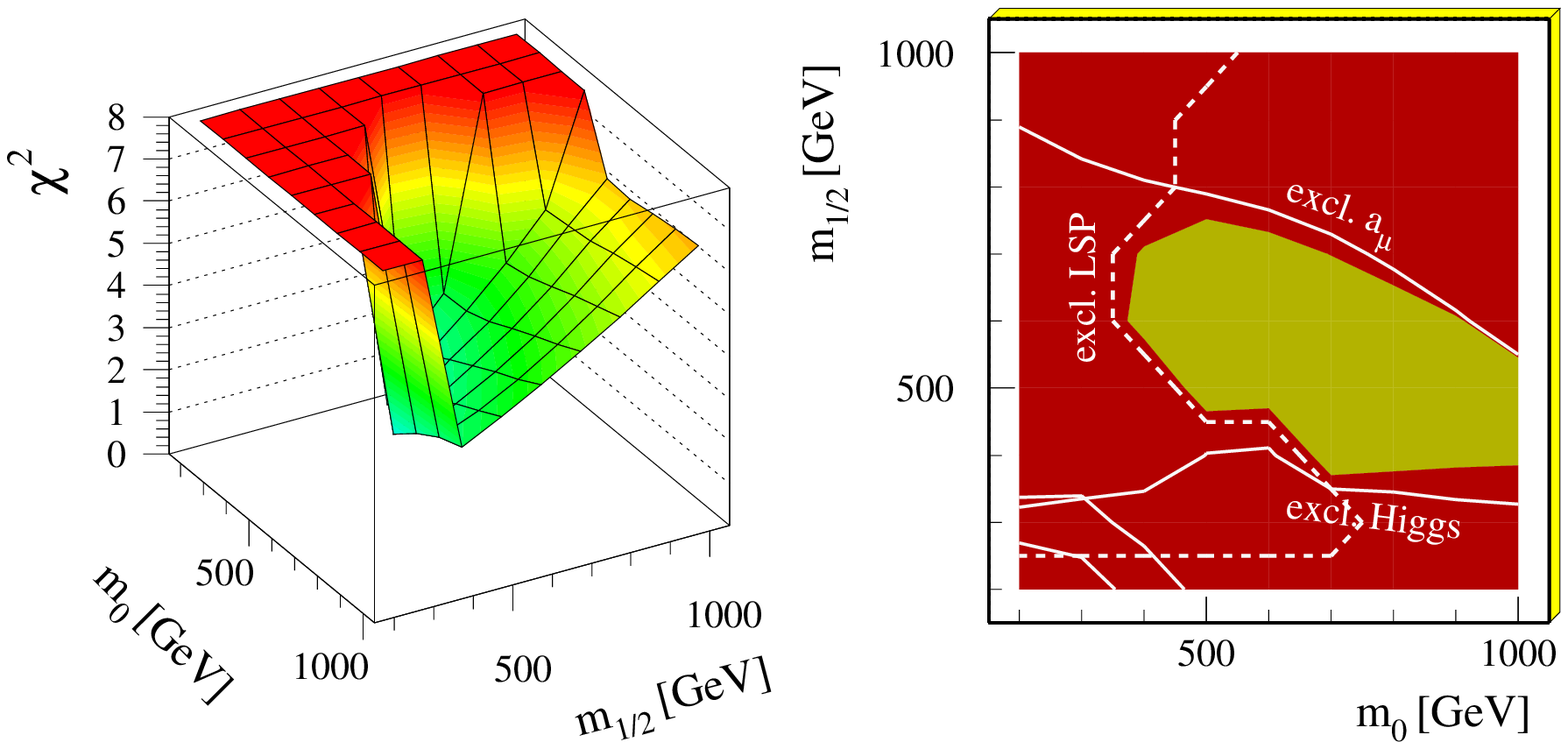,width=0.85\textwidth}
\caption[]{\label{f7}The total $\chi^2$ and the allowed regions in the 
parameter space for $\mu>0$ and $\tb =20$(top) and 50(bottom), with $A_0$
free, as in Fig. \ref{f6} (bottom).}
\end{center}
\end{figure}
We found that the allowed area of overlap
between \besg and $a_\mu$ can be increased considerably
for positive values of the common trilinear coupling $A_0$ at the GUT scale.
For  $A_0>0$ the present Higgs limit
becomes more stringent than for the no-scale models with
$A_0=0$,
as will be shown.

\section{{\boldmath $a_\mu$} and {\boldmath $b\to X_s\gamma$} in the CMSSM}

The contribution to the anomalous magnetic moment of the muon from SUSY 
particles are similar to  that of the weak interactions after replacing  
the vector bosons by charginos and neutralinos. 
The total contribution to $a_\mu$ can be approximated by~\cite{CM}
\begin{equation}  
|a_\mu^{SUSY}| \simeq \frac{\alpha(M_Z)}{8\pi \sin^2\theta_W}
\frac{m_\mu^2}{m_{SUSY}^2}\tan\beta\left(1-\frac{4\alpha}{\pi}
\ln\frac{m_{SUSY}}{m_\mu}\right) \simeq 140 \cdot 10^{-11} 
 \left(\frac{100 \ GeV}{m_{SUSY}}\right)^2 \tb, \label{1}
\end{equation}where $m_\mu$ is the muon mass, $m_{SUSY}$ is an average mass of supersymmetric
particles in the loop (essentially the chargino mass).
In our calculations we use the complete one-loop 
SUSY contributions from~\cite{CN} with zero phase factors and
the additional logarithmic
suppression factor as in eq.(\ref{1}). The calculated value of $a_\mu$ is
shown in Fig. \ref{f2}
as function of $\tb$. Clearly, it is approximately proportional to $\tb$ and
its sign 
depends on the sign of $\mu_0$\footnote{Our sign conventions are as in
Ref.~\cite{HK}.}.
Only positive values of $\mu_0$ are
allowed for the positive deviation from the SM and in addition the sparticles
have to be rather light.  However,  light sparticles
contribute also substantially to the \besg decay rate. In the past this posed a conflict. 
However, if one uses in the \besg calculations the running mass for the
charm quark, as suggested recently by Gambino and Misiak, the SM prediction is 
increased by 11\%. In this case the newest world average on \besg is
$1.7 \sigma$ below the SM, as mentioned in the introduction.
Such a deviation is most easily obtained for
large \tb and not too heavy sparticles, as shown in Fig. \ref{f3}. In the upper part
the scale uncertainty of the low energy scale $\mu_b$ is displayed by
the width of the theoretical curves, while in
the lower part the dependence on the trilinear coupling $A_0$ is shown.
The scale $\mu_b$ was varied between 0.5$m_b$ and 2$m_b$.
For \tb$\approx 40$ only positive values of the Higgsmixing parameter at the GUT scale 
$\mu_0$ are allowed
in agreement with the preferred sign of $\mu_0$ by the anomalous magnetic moment.
For intermediate sparticle masses and $\mu_0>0$ large values of $A_0$ and small
values of the low energy scale ($\mu_b\approx 0.5m_b$) bring the calculated values
of \besg closest to the data, as can be seen from the left hand side of Fig. \ref{f3}.
Note that for heavy sparticles (right hand side of Fig. \ref{f3}) the effect of the 
trilinear coupling is small, because the stop mixing is small, if the left and
right handed stops are much heavier than the top mass.\\
Fig. \ref{f4} shows the values of \besg and $a_\mu^{SUSY}$ as function of $m_0$ and
$m_{1/2}$ for \tb=35. For \besg the ratio $m_c(\mu)/m_b^{pole}=0.22$ was used, while
for the NLO QCD contributions the formulae from Ref. \cite{DGG} were used.
The calculated values have to be compared with the experimental values 
$BR(b\to X_s\gamma) = (2.96\pm 0.46) \times 10^{-4}$ ~\cite{CLEO}-\cite{BELLE} 
and  $\Delta a_\mu=(43\pm 16)\cdot 10^{-10}$ ~\cite{BNL},
which shows once more that $b\to X_s\gamma$ and $a_\mu^{SUSY}$ prefer a 
relatively light supersymmetric spectrum.

To find out the allowed regions in the parameter space of the CMSSM,
we  fitted both the $b\to X_s\gamma$ and $a_\mu$
data simultaneously. The fit includes the following constraints:
i) the unification of the gauge couplings, ii) radiative elctroweak symmetry
breaking, iii) the masses of the third generation particles, 
iv) $b\to X_s\gamma$ and $\Delta a_\mu$, v) experimental
limits on the SUSY masses, 
vi) the lightest superparticle (LSP) has to be neutral
to be a viable candidate for dark matter.  
We do not impose $b-\tau$ unification, since it prefers 
$\mu_0 <0$, as shown in Fig. \ref{f1}, while $\Delta a_\mu$ requires $\mu_0 >0$,
as shown in Fig. \ref{f2}.   
Yukawa unification for $\mu_0 >0$  can only be obtained by relaxed unification 
of the gauge couplings and  non-universality of the  soft terms in the 
Higgs sector~\cite{R}.

The $\chi^2$ contributions of $  b\to X_s\gamma$ and the anomalous
magnetic moment $a_\mu$ in the global fit are shown in Fig. \ref{f5} for
$A_0=0$  and $\tb=35$.
As expected, the $\chi^2$ contribution from $b\to X_s\gamma$ is smallest
for heavy sparticles, if \besg is calculated with $m_c/m_b=0.29$, while the
minimum $\chi^2$ is obtained for intermediate sparticles, if
$m_c/m_b=0.22$ is used. With the newly 
calculated \besg values, one can see, that \besg and $a_{\mu}$ prefer
a similar region of the $m_0,m_{1/2}$ plane.
Fig. \ref{f6} shows the combined $\chi^2$ contributions from \besg
and $a_\mu^{SUSY}$ in the $m_0$, $m_{1/2}$ plane, both in 3D and 2D,
for $A_0=0$ (top) and $A_0$ free (bottom).
In the latter case the lower $2\sigma$ contour from
$b\to X_s\gamma$ moves to the lower left corner, but for the preferred
value $A_0\approx 3m_0$, which is the maximum allowed value
in the fit in order to avoid negative stop- or Higgs  masses and
colour breaking minima, the Higgs bound moves up considerably.
The total allowed region is similar in both cases, as shown
by the light shaded areas in the contour plots. 
The $2\sigma$ contours from the individual contributions are 
in good agreement with previous calculations~\cite{FM,E}, 
but in these paper a simple scan over the parameter space 
was performed without calculating the combined probability.
In addition, $A_0=0$ was assumed.

We repeated the fits for $\tb=20$ and 50, as shown in
Fig. \ref{f7}. For smaller values of $\tb$ the allowed
region decreases, since $a_\mu$ becomes too small.
 At larger $\tb$ values the region allowed by $a_\mu$ and \besg increases
towards heavier sparticles, as expected from Eq. \ref{1}, but 
it is cut by the region where the charged stau lepton becomes
the Lightest Supersymmetric Particle (LSP), which is assumed to be stable and
should be neutral. A charged stable LSP
would have been observed by its electromagnetic interactions
after being produced in the beginning of the
universe. Furthermore, it would not be a candidate for dark matter.
The increase of the LSP-excluded area is due to the larger 
mixing term between the left- and right handed
staus at larger $\tb$. 

\vspace*{2mm}
We conclude that the $a_\mu$ measurement strongly
restricts the allowed region of the parameter space in the CMSSM,
since it excludes the $\mu_0<$ solution, which was the preferred one from
$b-\tau$ Yukawa unification. In addition, it prefers large $\tb$ with
relatively light sparticles, if the present deviation from the SM of $2.6\sigma$ persists.

 At large $\tb$ a global fit including both $b\to X_s\gamma$ and
$a_\mu$ as well as the present Higgs limit of 113.5 GeV
 leaves a quite large region in the CMSSM parameter space.
Here we left the trilinear coupling to be a
free  parameter, which  affects both the Higgs limit constraint 
 and  the \besg constraint, but in opposite ways, so that the preferred region
 is similar for the no-scale models with $A_0=0$ and models which leave
 $A_0$ free.

The 95\% lower limit on $m_{1/2}$ is $ 300$ GeV 
(see Figs. \ref{f6}+\ref{f7}), which implies that the lightest chargino
(neutralino) is above 240(120) GeV. The 95\% upper limit on $m_{1/2}$ is determined
by the lower limit on $a_\mu^{SUSY}$ and therefor depends on \tb (see Fig. \ref{f2}).
For \tb=35(50) one finds $m_{1/2}\leq 610(720)$ GeV, which implies that the lightest
chargino is below 500(590) GeV and the lightest neutralino is below 260(310) GeV.
\section*{Acknowledgements}

D.K. would like to thank the Heisenberg-Landau Programme,
RFBR grant \# 99-02-16650 and DFG grant \# 436/RUS/113/626 for
financial support and  the Karlsruhe University for hospitality
during completion of this work.

\end{document}